\documentclass[
]{ceurart}

\usepackage{listings}
\begin{document}

\copyrightyear{2023}
\copyrightclause{Copyright for this paper by its authors.
  Use permitted under Creative Commons License Attribution 4.0
  International (CC BY 4.0).}

\conference{}

\title{Contextualizing Problems to Student Interests at Scale in Intelligent Tutoring System Using Large Language Models}


\author[1]{Gautam Yadav}[%
orcid=0000-0002-9247-1920 ,
email=gyadav@andrew.cmu.edu,
]

\author[1]{Ying-Jui Tseng}[%
orcid=0009-0006-1801-6061,
email=yingjuit@andrew.cmu.edu,
]
\author[1]{Xiaolin Ni}[%
orcid=0009-0009-2660-9539,
email=xiaolinn@andrew.cmu.edu,
]

\address[1]{Carnegie Mellon University,
  5000 Forbes Ave Pittsburgh PA 15213, United States}

\begin{abstract}
  Contextualizing problems to align with student interests can significantly improve learning outcomes. However, this task often presents scalability challenges due to resource and time constraints. Recent advancements in Large Language Models (LLMs) like GPT-4~\cite{openai2023gpt4}offer potential solutions to these issues. This study explores the ability of GPT-4 in the contextualization of problems within CTAT \cite{aleven2006cognitive}, an intelligent tutoring system, aiming to increase student engagement and enhance learning outcomes. Through iterative prompt engineering, we achieved meaningful contextualization that preserved the difficulty and original intent of the problem, thereby not altering values or overcomplicating the questions. While our research highlights the potential of LLMs in educational settings, we acknowledge current limitations, particularly with geometry problems, and emphasize the need for ongoing evaluation and research. Future work includes systematic studies to measure the impact of this tool on students' learning outcomes and enhancements to handle a broader range of problems.
\end{abstract}


\maketitle

\section{Introduction}

Research has demonstrated that integrating problem contextualization with student interests can significantly enhance learning outcomes in algebra, resulting in increased proficiency in problem-solving, improved accuracy, and the ability to transfer to future learning \cite{walkington2013using}. 

Teachers, who intimately comprehend their students' interests, often find the task of contextualizing problems according to these interests challenging, since the scalability of such task is often met with resource and time constraints. However, recent developments in Large Language Models (LLMs) may provide an opportunity to lessen the strains associated with the personalization of learning context for students.

This research aims to explore the capability of LLMs in contextualizing problems to align with student interests at a large scale within CTAT \cite{aleven2006cognitive}, an intelligent tutoring system. In this study, we perform experiments using one of the most advanced LLMs currently accessible, the GPT-4 \cite{openai2023gpt4}, obtained via the OpenAI API. Our hypothesis suggests that the application of LLMs for problem contextualization, based on student interests, could result in increased student engagement and enhanced learning outcomes.

\section{Prior Work}

\subsection{Context Personalization}

The groundbreaking works of Walkington \cite{walkington2013using} introduced the concept of contextualizing algebraic questions based on students' interests. This innovative methodology, featuring student-created "algebra stories," aimed to boost engagement, cultivate ownership, and enhance understanding of algebraic principles. Interest has been identified as a pivotal factor in learning, impacting attention, persistence, and motivation. Personalized learning that mirrors individual interests has demonstrated a capacity to elicit positive emotional responses, enhance appreciation for instructional content, and leverage existing knowledge. \cite{hidi2006four, goldstone2005transfer}

The efficacy of context personalization was investigated using both qualitative and quantitative research methods, demonstrating a positive association between these 'algebra stories' and improved student engagement and performance. Despite possible implementation obstacles due to the diversity of learners' interests, the use of digital tools has been proposed as a facilitative means for this personalization process. In its totality, contextual personalization has the potential to enhance learning effectiveness and accuracy, decrease the practice required for mastery, and foster transferable skills applicable to various scenarios. 

\subsection{Mass Production in Intelligent Tutoring Systems}

Mass Production in Intelligent Tutoring Systems (ITS) is a technique that enables authors to parameterize previously authored problem-specific content, which can then be instantiated to suit a multitude of different problems. This technique essentially permits authors to manually generalize Example-Tracing expert models (known as behavior graphs) to accommodate all problems that share isomorphic solution structures.

The application of mass production in ITS offers significant value, as it facilitates the creation of a vast array of distinct problems using the same underlying structure. This contributes to mastery learning, allowing learners to practice similar problems in various contexts, ultimately strengthening their grasp of the subject \cite{maclellan2022domain}.

In our approach, we utilized this principle, where only the contextual 'cover stories' were varied for the problem within the CTAT platform. This delivers similar problem-solving opportunities to students, yet personalizes these scenarios to align with their individual interests. The implications of this mass production approach based on interests are manifold; it can potentially increase student engagement, improve problem-solving abilities, and promote a better understanding of the subject matter.

\subsection{Instruction Generation with Large Language Models}

Previous research involving large language models has explored their application in educational settings, such as the use of models like GPT for generating questions or providing hints/explanations to students \cite{elkins2023useful}. Empirical evaluations of these applications and their impact on student outcomes suggested that students perform better on content generated by these models compared to human-generated content~\cite{prihar2023comparing,pardos2023learning}.

It is evident that large language models hold great potential in enhancing learning experiences, making them a promising tool for future educational endeavors. However, as our work proposes, a step further in personalized learning can be taken by leveraging these models in a more context-aware manner, which could further improve student engagement and outcomes.

\section{System Design}

\subsection{Iterative Prompt Engineering in GPT-4}

Our iterative prompt engineering was executed over four different problem sets in TutorShop~\cite{aleven2015beginning}, with the objective of achieving meaningful contextualization aligned with diverse interests. A few-shot learning approach was used, drawing on examples from a range of contexts and interests to foster coverage and generalization. Throughout this process, we gradually refined and added rules based on testing until our output could aptly accommodate novel interests such as TikTok and NBA, as demonstrated in Tables 1 and 2.

\begin{table*}
  \caption{Contextualized Algebra Problem with the cover story based on interests Tiktok and NBA}
  \label{tab:freq}
  \begin{tabular}{p{0.2\textwidth}p{0.7\textwidth}}
    \toprule
    Interest & Question \\
    \midrule
    Original & Danny and the Algebraics are recording their first demo album. They plan to send 15 CDs to record companies to try to land a recording contract. They are wondering how many additional CDs they should make. They have a fixed budget of \$1000 and it costs \$2.50 to make a CD. They use this formula to see how much they will have left if they make a different number of CDs.

    Let $C$ be the number of CDs they make after the initial 15 CDs.

    The amount of money they will have left = 1000 - 2.50$(C+15)$

    \begin{enumerate}
      \item How much money is left if they make 85 additional CDs?
      \item How much money is left if they make 125 additional CDs?
      \item How much money is left if they make 250 additional CDs?
      \item How much money is left if they make 385 additional CDs?
    \end{enumerate} \\
    TikTok & An upcoming TikTok creator, Danny and his creative team, are planning to promote their first viral video. They plan to run a paid advertising campaign to 15 prominent TikTok influencers to try to gain more followers. They are wondering how many more influencers they should reach out to. They have a fixed budget of \$1000 and it costs \$2.50 to run an ad per influencer. They use this formula to see how much they will have left if they promote to a different number of influencers.
    Let $I$ = the number of influencers they target after the initial 15 influencers.
    The amount of money they will have left = 1000-2.50$(I+15)$
    \begin{enumerate}
    \item How much money is left if they reach out to 85 additional influencers?
    \item How much money is left if they reach out to 125 additional influencers?
    \item How much money is left if they reach out to 250 additional influencers?
    \item How much money is left if they reach out to 385 additional influencers?
    \end{enumerate} \\
    NBA & The Lakers are planning their season's training sessions. They have a budget of \$1000 to buy basketballs for practice. Each basketball costs \$2.50. They are wondering how many additional basketballs they should buy. They already plan to buy 15 basketballs to start off. They use this formula to see how much they will have left if they buy a different number of basketballs.
    Let $B$ = the number of basketballs they buy after the initial 15 basketballs.
    The amount of money they will have left = 1000-2.50$(B+15)$
    \begin{enumerate}
    \item How much money is left if they buy 85 additional basketballs?
    \item How much money is left if they buy 125 additional basketballs?
    \item How much money is left if they buy 250 additional basketballs?
    \item How much money is left if they buy 385 additional basketballs?
    \end{enumerate} \\
        
  \bottomrule
\end{tabular}
\end{table*}

\begin{table*}
  \caption{Contextualized Algebra Problem without cover story based on interests Tiktok and NBA}
  \label{tab:freq}
  \begin{tabular}{p{0.2\textwidth}p{0.7\textwidth}}
    \toprule
    Interest & Question \\
    \midrule
    Original & 2x + 3 = 15 \\
    TikTok & In the realm of TikTok, you're working on a new viral challenge. You have found a way to get 2 new followers every time you post a video. Additionally, you already have 3 followers to start with. Your aim is to have 15 followers.
    Create an equation to show how many TikTok videos 'x' you need to post to reach your follower goal. \\
    NBA & During an NBA game, a player earns points for their team by scoring baskets. Each 2-point field goal adds 2 points and every free throw adds a single point to the team's total. Imagine a situation where a player, LeBron, makes a certain number of 2-point field goals and 3 successful free throws, resulting in 15 points for his team.
    Write an equation that would help determine the number of 2-point field goals LeBron made. Use 'x' to denote the number of 2-point field goals. \\
  \bottomrule
\end{tabular}
\end{table*}

\subsubsection{Prompt}
We used the following prompt:
\begin{itemize}
    \item Your task is to change context based on interest for a problem, for example:
    \begin{itemize}
    \item Input Problem 1:
    
Chaz and Nikki are standing in a long line to buy rock concert tickets. Nikki is 8 feet ahead of Chaz in the line. Let's compare Chaz's distance to Nikki's distance from the front of the line.

When Nikki is 20 feet from the front of the line, how far away is Chaz?

When Nikki is 16 feet from the front of the line, how far away is Chaz?

In the row labeled "Expression", define a variable for Nikki's distance and use that variable to write an expression that will allow you to calculate Chaz's distance. \\

Output Problem 1 based on interest "Video Games":

In a video game, two players, Mario and Luigi, are standing at different points in a level. Luigi is 8 units ahead of Mario in the game. Let's compare Mario's distance to Luigi's distance from the level's end.

When Luigi is 20 units from the end of the level, how far away is Mario?

When Luigi is 16 units from the end of the level, how far away is Mario?

In the row labeled "Expression", define a variable for Mario's distance and use that variable to write an expression that will allow you to calculate Luigi's distance. \\

Output Problem 1 based on interest "basketball":

During a basketball game, two players, Jordan and Kobe, are standing at different positions on the court. Jordan is 12 feet ahead of Kobe on the court. Let's compare Jordan's distance to Kobe's distance from the basket.

When Kobe is 20 feet away from the basket, how far away is Jordan from the basket?

When Kobe is 16 feet away from the basket, how far away is Jordan from the basket?

In the row labeled "Expression", define a variable for Kobe's distance and use that variable to write an expression that will allow you to calculate Jordan's distance. \\

\item Input Problem 2:

You are a product inspector for a company that produces light bulbs. You find that two out of every 300 bulbs are defective: they don’t work properly. \\

Output Problem 2 based on interest "World of Warcraft":

You enjoy playing World of Warcraft on your computer. You notice that two out of every 300 times you defeat a monster, the monster has an epic item: a treasure that you want to collect. \\

\item Input Problem 3:

y = 80 - 6x

If x = 10, what is y?

If x = 7, what is y?

If y = 8, what is x?

Write a story that could go along with the equation y = 80 - 6x. \\

Output Problem 3 based on interest "Video Games":

You are playing your favorite war game on the Xbox 360. When you started playing today, there were 80 enemies left in the locust horde. You kill an average of 6 enemies every minute.

(a) How many enemies are left after 10 minutes?

(b) How many enemies are left after 7 minutes?

(c) Write an algebra rule that represents this situation using symbols.

(d) If there are only 8 enemies left, how long have you been playing today?
\end{itemize}
Now give output for 
\begin{itemize}
\item input problem: 2x+3=15
\item Interest: [The interest that the problem needs to be contextualized for.]
\end{itemize}
Some rules to follow:
\begin{enumerate}
\item don't change values
\item we want to have deeper contextualization not surface details based on Using Adaptive Learning Technologies to Personalize Instruction to Student Interests: The Impact of Relevant Contexts on Performance and Learning Outcomes
\item output question should ask same thing as input question, don't ask any additional question or complicate the info by adding unnecessary details
\end{enumerate}
\end{itemize}

This strict adherence to rules ensures that we maintain consistency in problem difficulty and preserve the problem's original intent. This methodology respects the principle of not altering values or over-complicating the question by adding unnecessary details as observed in our earlier iterations.

\subsection{CTAT Implementation}
In this section, we propose a novel interaction design for contextualizing problems in Intelligent Tutoring Systems using CTAT and GPT-4 that emphasizes problem-authoring control. Teachers or instructional designers could contextualize existing problems simply by adding interest in the “Contextualized by Interest”  tab in the Mass Production feature (Figure 1). After the user click the contextualize problem, the system will use GPT-4 and the prompt we mentioned in the prompt engineering section to generate variations of the problem for each interest. They can also preview and edit the contextualized result in the student-facing interface on the right panel to make sure whether they are satisfied with the generation result (Figure 2).

\begin{figure}
  \centering
  \includegraphics[width=\linewidth]{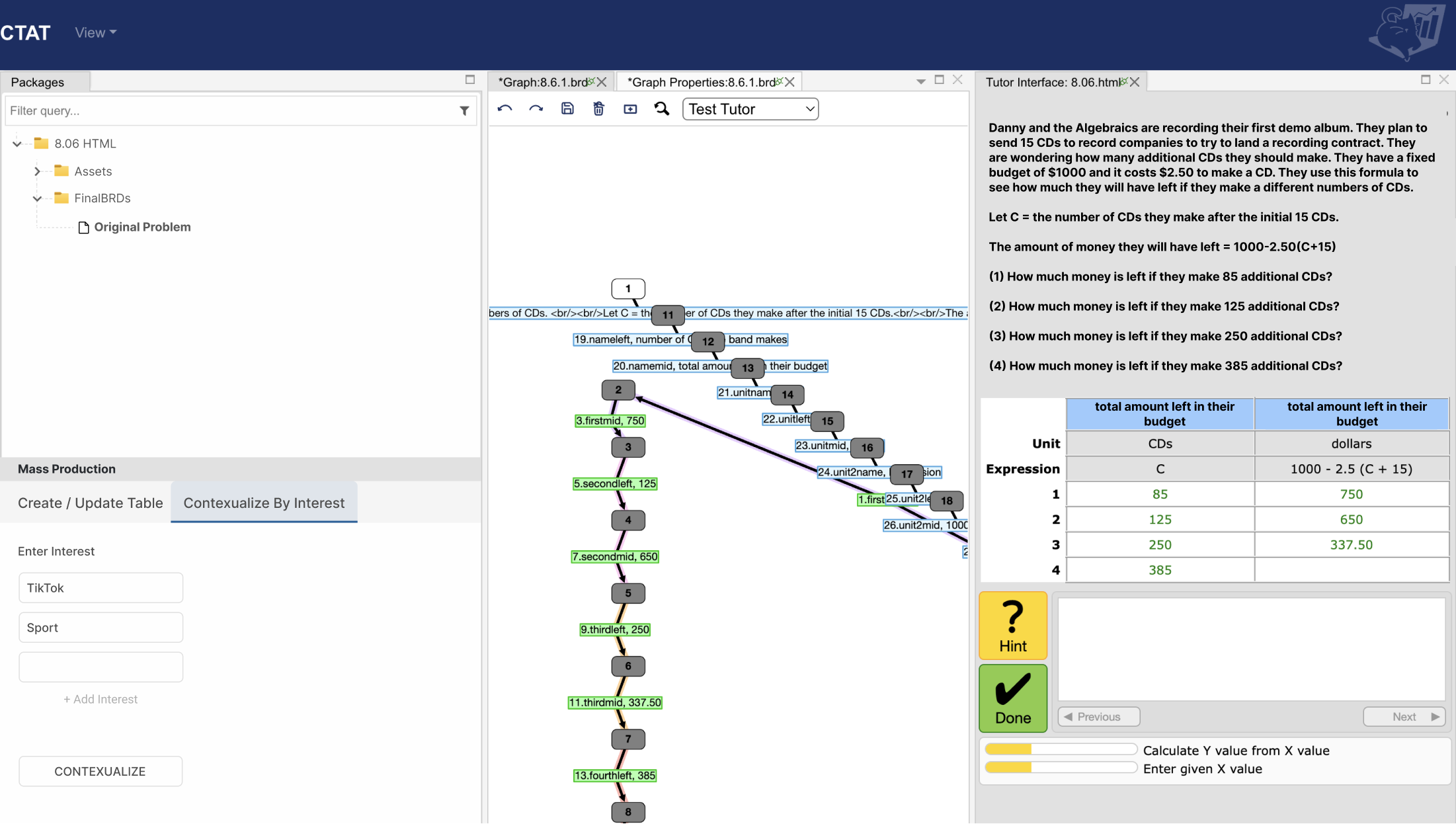}
  \caption{ User can enter or add interests in the  “Contexualize By Interest” tab
    }
    \label{fig:ui}
\end{figure}

\begin{figure}
  \centering
  \includegraphics[width=\linewidth]{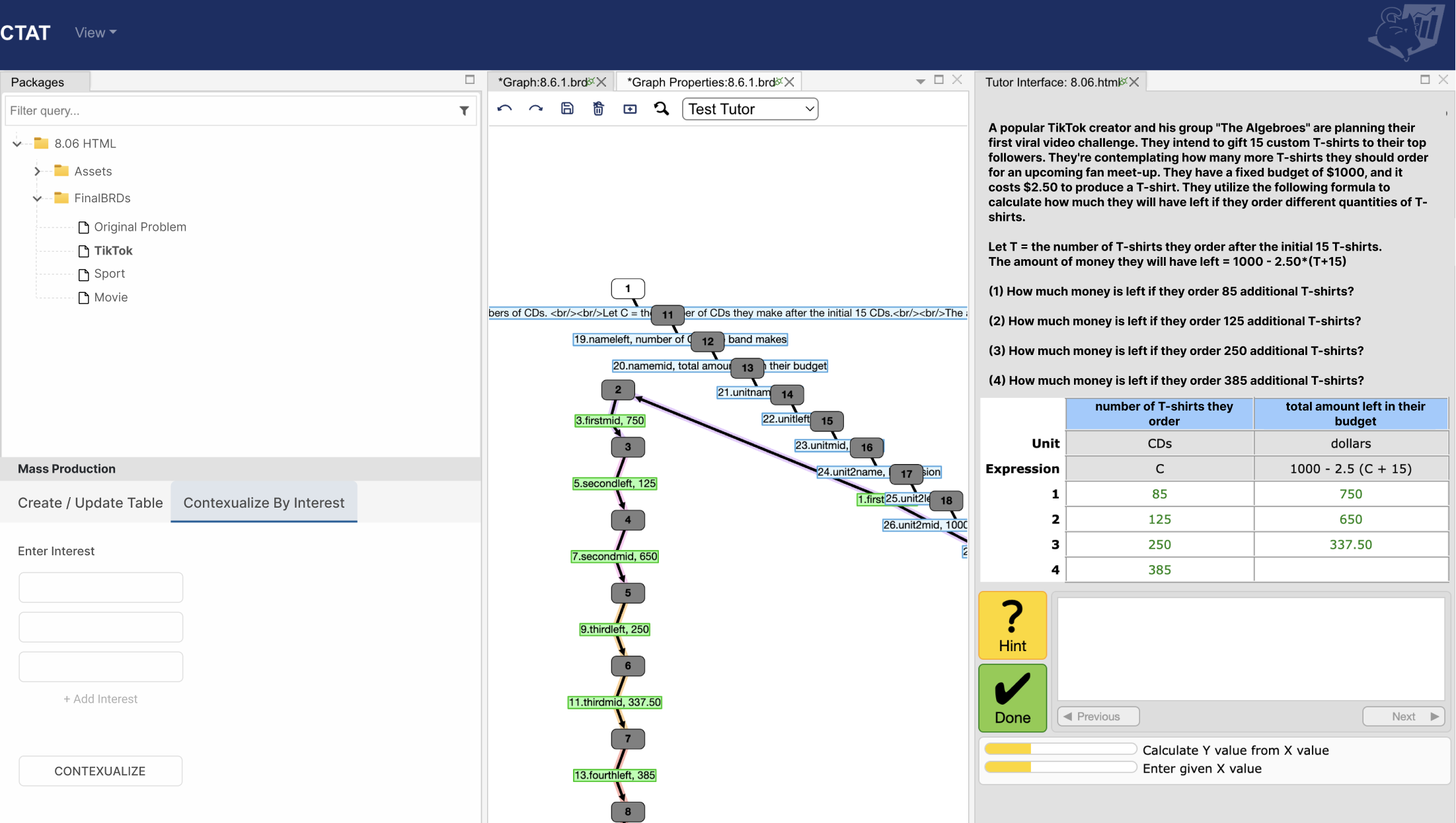}
  \caption{After user click the “CONTEXUALIZE” button, they can preview and edit the contexualized problems. For example, this figure show how the system contexualized the original algebra problem based on the “TikTok” interest.
    }
    \label{fig:ui2}
\end{figure}

\section{Future Work and Limitations}
While this work is firmly grounded in existing pedagogical and technological research, it is imperative that ongoing evaluation continues to ensure its effective application within real-world educational environments. We plan to conduct systematic studies to measure the impact of this tool on students' learning outcomes. This encompasses improvements in the initial accuracy of responses, an enhancement in learning efficiency, and an accelerated pace toward proficiency.

However, certain limitations in the current model require attention. While it excels in solving algebraic equations, it needs help with geometric problems, especially those involving graphs, tables, or diagrammatic components. Its existing capabilities of GPT limit its ability to create images that align with the problem text, accurately representing variable relationships. Specifically, the system fails to produce suitable diagrams for linear algebra questions requiring visual components, which is essential for testing students' comprehension of the underlying concepts.

\begin{acknowledgments}
  We extend our sincere gratitude to Prof. Vincent Aleven, whose expert guidance was indispensable to the success of this research. His profound wisdom and unwavering support enriched this work immeasurably. 
\end{acknowledgments}

\bibliography{REFERENCES}

\begin{thebibliography}{10}
\expandafter\ifx\csname natexlab\endcsname\relax\def\natexlab#1{#1}\fi
\providecommand{\url}[1]{\texttt{#1}}
\providecommand{\href}[2]{#2}
\providecommand{\path}[1]{#1}
\providecommand{\DOIprefix}{doi:}
\providecommand{\ArXivprefix}{arXiv:}
\providecommand{\URLprefix}{URL: }
\providecommand{\Pubmedprefix}{pmid:}
\providecommand{\doi}[1]{\href{http://dx.doi.org/#1}{\path{#1}}}
\providecommand{\Pubmed}[1]{\href{pmid:#1}{\path{#1}}}
\providecommand{\bibinfo}[2]{#2}
\ifx\xfnm\relax \def\xfnm[#1]{\unskip,\space#1}\fi
\bibitem[{OpenAI(2023)}]{openai2023gpt4}
\bibinfo{author}{OpenAI}, \bibinfo{title}{Gpt-4 technical report},
  \bibinfo{year}{2023}. \href{http://arxiv.org/abs/2303.08774}{{\tt
  arXiv:2303.08774}}.
\bibitem[{Aleven et~al.(2006)Aleven, McLaren, Sewall, and
  Koedinger}]{aleven2006cognitive}
\bibinfo{author}{V.~Aleven}, \bibinfo{author}{B.~M. McLaren},
  \bibinfo{author}{J.~Sewall}, \bibinfo{author}{K.~R. Koedinger},
\newblock \bibinfo{title}{The cognitive tutor authoring tools (ctat):
  Preliminary evaluation of efficiency gains},
\newblock in: \bibinfo{booktitle}{Intelligent Tutoring Systems: 8th
  International Conference, ITS 2006, Jhongli, Taiwan, June 26-30, 2006.
  Proceedings 8}, \bibinfo{organization}{Springer}, \bibinfo{year}{2006}, pp.
  \bibinfo{pages}{61--70}.
\bibitem[{Walkington(2013)}]{walkington2013using}
\bibinfo{author}{C.~A. Walkington},
\newblock \bibinfo{title}{Using adaptive learning technologies to personalize
  instruction to student interests: The impact of relevant contexts on
  performance and learning outcomes.},
\newblock \bibinfo{journal}{Journal of educational psychology}
  \bibinfo{volume}{105} (\bibinfo{year}{2013}) \bibinfo{pages}{932}.
\bibitem[{Hidi and Renninger(2006)}]{hidi2006four}
\bibinfo{author}{S.~Hidi}, \bibinfo{author}{K.~A. Renninger},
\newblock \bibinfo{title}{The four-phase model of interest development},
\newblock \bibinfo{journal}{Educational psychologist} \bibinfo{volume}{41}
  (\bibinfo{year}{2006}) \bibinfo{pages}{111--127}.
\bibitem[{Goldstone and Son(2005)}]{goldstone2005transfer}
\bibinfo{author}{R.~L. Goldstone}, \bibinfo{author}{J.~Y. Son},
\newblock \bibinfo{title}{The transfer of scientific principles using concrete
  and idealized simulations},
\newblock \bibinfo{journal}{The Journal of the learning sciences}
  \bibinfo{volume}{14} (\bibinfo{year}{2005}) \bibinfo{pages}{69--110}.
\bibitem[{MacLellan and Koedinger(2022)}]{maclellan2022domain}
\bibinfo{author}{C.~J. MacLellan}, \bibinfo{author}{K.~R. Koedinger},
\newblock \bibinfo{title}{Domain-general tutor authoring with apprentice
  learner models},
\newblock \bibinfo{journal}{International Journal of Artificial Intelligence in
  Education}  (\bibinfo{year}{2022}) \bibinfo{pages}{1--42}.
\bibitem[{Elkins et~al.(2023)Elkins, Kochmar, Cheung, and
  Serban}]{elkins2023useful}
\bibinfo{author}{S.~Elkins}, \bibinfo{author}{E.~Kochmar},
  \bibinfo{author}{J.~C. Cheung}, \bibinfo{author}{I.~Serban},
\newblock \bibinfo{title}{How useful are educational questions generated by
  large language models?},
\newblock \bibinfo{journal}{arXiv preprint arXiv:2304.06638}
  (\bibinfo{year}{2023}).
\bibitem[{Prihar et~al.(2023)Prihar, Lee, Hopman, Kalai, Vempala, Wang,
  Wickline, and Heffernan}]{prihar2023comparing}
\bibinfo{author}{E.~Prihar}, \bibinfo{author}{M.~Lee},
  \bibinfo{author}{M.~Hopman}, \bibinfo{author}{A.~Kalai},
  \bibinfo{author}{S.~Vempala}, \bibinfo{author}{A.~Wang},
  \bibinfo{author}{G.~Wickline}, \bibinfo{author}{N.~Heffernan},
\newblock \bibinfo{title}{Comparing different approaches to generating
  mathematics explanations using large language models},
\newblock in: \bibinfo{booktitle}{Proceedings of the AIED2023 Conference},
  \bibinfo{year}{2023}. \bibinfo{note}{To be published}.
\bibitem[{Pardos and Bhandari(2023)}]{pardos2023learning}
\bibinfo{author}{Z.~A. Pardos}, \bibinfo{author}{S.~Bhandari},
\newblock \bibinfo{title}{Learning gain differences between chatgpt and human
  tutor generated algebra hints},
\newblock \bibinfo{journal}{arXiv preprint arXiv:2302.06871}
  (\bibinfo{year}{2023}).
\bibitem[{Aleven et~al.(2015)Aleven, Sewall, Popescu, Xhakaj, Chand, Baker,
  Wang, Siemens, Ros{\'e}, and Gasevic}]{aleven2015beginning}
\bibinfo{author}{V.~Aleven}, \bibinfo{author}{J.~Sewall},
  \bibinfo{author}{O.~Popescu}, \bibinfo{author}{F.~Xhakaj},
  \bibinfo{author}{D.~Chand}, \bibinfo{author}{R.~Baker},
  \bibinfo{author}{Y.~Wang}, \bibinfo{author}{G.~Siemens},
  \bibinfo{author}{C.~Ros{\'e}}, \bibinfo{author}{D.~Gasevic},
\newblock \bibinfo{title}{The beginning of a beautiful friendship? intelligent
  tutoring systems and moocs},
\newblock in: \bibinfo{booktitle}{Artificial Intelligence in Education: 17th
  International Conference, AIED 2015, Madrid, Spain, June 22-26, 2015.
  Proceedings 17}, \bibinfo{organization}{Springer}, \bibinfo{year}{2015}, pp.
  \bibinfo{pages}{525--528}.

\end{thebibliography}

\appendix

\end{document}